\documentstyle[12pt]{article}
\title{\bf \Huge Generalized Laws of Black Hole Thermodynamics and  
                 Quantum Conservation Laws on Hawking Radiation Process } 
\author{S. Q. Wu\thanks{E-mail: sqwu@iopp.ccnu.edu.cn} 
    and X. Cai\thanks{E-mail: xcai@wuhan.cngb.com} \\   
  \footnotesize \sl Institute of Particle Physics, Hua-Zhong 
                    Normal University, Wuhan, 430079, China }

\date{\today}
\begin{document}
\maketitle
\baselineskip 25pt
\begin{quote}

Four classical laws of black hole thermodynamics are extended from 
exterior (event) horizon to interior (Cauchy) horizon. Especially, the 
first law of classical thermodynamics for Kerr-Newman black hole (KNBH) 
is generalized to those in quantum form. Then five quantum conservation laws 
on the KNBH evaporation effect are derived in virtue of thermodynamical 
equilibrium conditions. As a by-product, Bekenstein-Hawking's relation 
$ S=A/4 $ is exactly recovered. 
 
PACS number(s): 04.70.Dy, 97.60.Lf               
\end{quote}
\vskip 0.5cm

Due to the celebrated works of Hawking [1] and Bekenstein [2], black
holes are demonstrated to be thermodynamic objects endowed with a 
temperature and an entropy. This put the first laws of black hole
thermodynamics on a solid fundament [3]. Despite considerable effort 
[4] about the quantum [5], dynamic [6], or statistical [7] origin of
black hole thermodynamics, however, the exact source and mechanism of 
the Benkenstein-Hawking entropy remain unclear [8].

In this paper, we assume that the whole space-time of a rotating charged
black hole is described by Kerr-Newman metric and base our discussion on 
a sourceless charged massive scalar field on Kerr-Newman black hole (KNBH) 
background in the non-extreme case ($ 0 < \varepsilon=\sqrt{M^2-a^2-e^2} 
\leq M $). We propose that there should exist an inner thermal radiation
on the internal horizon and that a Kerr-Newman black hole have a pair of 
entropy accompanied by a pair of temperature. Under this assumption, we 
can extend Bardeen-Carter-Hawking's four laws of black hole thermodynamics
on exterior horizon to those on internal horizon. We also suggest a pair of
quantum first laws of thermodynamics. Then quantum conservation laws on
black hole radiation process are derived from the point of view of 
thermodynamical equilibrium [9]. Finally, the relation between classical 
entropy and quantum entropy of black hole is briefly discussed.

\noindent
{\large 1. Preliminary: Separation of Covariant Klein-Gordon Equation 
        on KNBH Background }

In the Boyer-Lindquist coordinates, Kerr-Newman metric is given by [10,11]:
\begin{equation}
 ds^2=-\frac{\Delta}{\Sigma}(dt-a\sin^2\theta d\varphi)^2+\frac{\sin^2
     \theta}{\Sigma}[adt-(r^2+a^2)d\varphi]^2+\Sigma(\frac{dr^2}
     {\Delta}+d\theta^2)
\end{equation}

\noindent
with $ \Delta=r^2-2Mr+a^2+e^2=(r-r_+)(r-r_-), \Sigma=r^2+a^2\cos^2\theta,
r_{\pm}=M\pm \sqrt{M^2-a^2-e^2} $, where mass $ M $, charge $ e $ and
specific angular momentum $ a=J/M $ being three parameters of KNBH. (In 
Planck unit system $ G=\hbar=c=k_B=1 $).
 
In the KNBH geometry, a complex scalar field $ \Phi $ with mass $ \mu $ 
and charge $ q $ satisfies the following covariant Klein-Gordon equation 
(KGE) [10,12]:
\begin{eqnarray}\nonumber
  \frac{-1}{\Delta}[(r^2+a^2)\partial_t+a\partial_{\varphi}+iqer]^2\Phi
  +\partial_r(\Delta\partial_r\Phi)-{\mu}^2\Sigma\Phi \\
  +(a\sin\theta\partial_t+\frac{1}{\sin\theta}\partial_{\varphi})^2\Phi+
  \frac{1}{\sin\theta}\partial_{\theta}(\sin\theta\partial_{\theta}\Phi)
  =0.
\end{eqnarray}

The wave function $ \Phi $ of the above equation has a solution of variables 
separable form $ \Phi(t,r,\theta,\varphi)=R(r)S(\theta)e^{i(m\varphi-\omega 
t)} $ [11], in which the separated radial and angular part of KGE can be given
as follows 
[12]:
\begin{eqnarray}
 \frac{1}{\sin\theta}\partial_{\theta}[\sin\theta\partial_{\theta}
 S(\theta)]+[\lambda-\frac{m^2}{\sin^2\theta}+({\mu}^2-{\omega}^2)a^2
 \sin^2\theta]S(\theta)=0, \\
 \partial_r[\Delta\partial_rR(r)]+[\frac{K^2(r)}{\Delta}-{\mu}^2(r^2
 +a^2)-\lambda+2ma\omega]R(r)=0, 
\end{eqnarray}

\noindent
here $ \lambda $ is a separation constant, and $ K(r)=\omega(r^2+a^2)
-qer-ma. $    

The general solutions to the angular equation of Eq.(3) are ordinary 
spheroidal angular wave functions [13] with spin-weight $ s=0 $, while 
the radial equation of Eq.(4) can be reduced to the following generalized 
spin-weighted spheroidal wave equation [14] of imaginary number order 
(see Eqs.(13) and (14) in Ref. [12]):
\begin{eqnarray}\nonumber
 \partial_r[(r-r_+)(r-r_-)\partial_rR(r)]+[k^2(r-r_+)(r-r_-)+2D(r-M)+ \\
 \frac{[A(r-M)+\varepsilon B]^2}{(r-r_+)(r-r_-)}
 +(2{\omega}^2-{\mu}^2)(2M^2-e^2)-2qeM\omega-\lambda]R(r)=0,
\end{eqnarray}

\noindent
where we have put 
$$ A=2M\omega-qe, \varepsilon B=\omega(2M^2-e^2)-qeM-ma,
 D=A\omega-M{\mu}^2, k^2={\omega}^2-{\mu}^2. $$
  
Introducing $ w_{\pm}=(B \pm A)/2 $ for simplicity, and making substitution 
$ R(r)=(r-r_+)^{i(B+A)/2}(r-r_-)^{i(B-A)/2}F(r) $, we can transform Eq.(5) 
for $ R(r) $ into a modified generalized spheroidal wave equation with an 
imaginary spin-weight $ iA $ for $ F(r) $ [12]:
\begin{eqnarray}\nonumber
 (r-r_+)(r-r_-)\partial_r^2F(r)+2[i\varepsilon A+(1+iB)(r-M)]
 \partial_rF(r)+[k^2(r-r_+)(r-r_-)   \\   
 +2D(r-M)+(2{\omega}^2-{\mu}^2)(2M^2-e^2)
 -2qeM\omega-\lambda+A^2-B^2+iB]F(r)=0.
\end{eqnarray} 

Eq.(6) has two regular singular points $ r=r_{\pm} $ whose indices are 
$ \rho_+=0, -2iw_+ $ and $ \rho_-=0, -2iw_- $ respectively. The general 
solutions to Eq.(6) have forms around regular singular points $ r_{\pm} $
\begin{eqnarray}
  F_+(r)&=&c_1f_1(A,B,D,k,r-r_+)+c_2(r-r_+)^{-2iw_+}g_1(A,B,D,k,r-r_+), \\
  F_-(r)&=&d_1f_2(A,B,D,k,r-r_-)+d_2(r-r_-)^{-2iw_-}g_2(A,B,D,k,r-r_-)
\end{eqnarray}

\noindent
where functions $ f_1, f_2 $ are first solutions to Eq.(6), while $ g_1, 
g_2 $ being linear independent second ones to it. They are four sets of 
orthonormal generalized spheroidal wave functions being regular at points 
$ r_+, r_- $ respectively. We choose such an eigenvalue $ \lambda $ that 
makes $ F_{\pm}(r) $ finite at $ r=r_{\pm} $ respectively, namely functions 
$ f_1, f_2 $ and $ g_1, g_2 $ are regular over their corresponding regions.
 
The physical domain for radial coordinate $ r $ is $ [0, \infty)=[0, r_-)
\cup (r_-, r_+) \cup (r_+, \infty) $, with region $ [0, {\mu}^2 /2) \cup 
[{\mu}^2 /2, {\mu}^2] \cup ({\mu}^2, \infty) $ for quadratic energy $ 
{\omega}^2 $. We can extend simultaneously both intervals on real axes 
for coordinate and that for energy to corresponding whole complex planes 
including real axes ($ -\infty, +\infty $). The extreme case $ \varepsilon
=0 $ and special cases $ \omega=\pm\mu /\sqrt{2} $, as well as $ \omega=
\pm\mu $ need to be carefully dealt with, but we don't discuss it here.
 
\noindent
{\large 2. Hawking Radiation: External or Internal? }

The exterior horizon and interior horizon, denoted by $ {\cal{H}}_{\pm} $, 
are located at points $ r_{\pm}=M\pm\varepsilon $. We shall consider a wave 
outgoing from  horizon $ {\cal{H}}_+ $ over intervals $ r_- < r < r_+ $ and 
$ r_+ < r < \infty $.

According to the method of Damour and Ruffini's [15], a correct outgoing 
wave $ \Phi^{\rm{out}}=\Phi^{\rm{out}}(t,r,\theta,\varphi) $ is an adequate 
superposition of functions $ \Phi_{r>r_+}^{\rm{out}} $ and $ \Phi_{r<r_+}^{
\rm{out}} $:
\begin{equation}
 \Phi^{\rm{out}}=C_+[\eta(r-r_+)\Phi_{r>r_+}^{\rm{out}}
  +\eta(r_+-r)\Phi_{r<r_+}^{\rm{out}}e^{2\pi w_+}]
\end{equation}

\noindent
where $ \eta(x) $ is conventional unit step function, with the outgoing wave 
components $ \Phi_{r>r_+}^{\rm{out}} $ and $ \Phi_{r<r_+}^{\rm{out}} $ being 
given by
\begin{eqnarray} 
 \Phi_{r>r_+}^{\rm{out}}(t,r,\theta,\varphi)
 \sim c_1(r-r_+)^{iw_+}(r-r_-)^{iw_-}f_{w_+,w_-}^{\ell}(k,r)
 S_{m,0}^{\ell}(ka,\theta)e^{i(m\varphi-\omega t)}, \\
 \Phi_{r<r_+}^{\rm{out}}(t,r,\theta,\varphi)
 \sim c_2(r-r_+)^{-iw_+}(r-r_-)^{iw_-}g_{w_+,w_-}^{\ell}(k,r)
 S_{m,0}^{\ell}(ka,\theta)e^{i(m\varphi-\omega t)},
\end{eqnarray}

\noindent
here regular function $ f_{w_+,w_-}^{\ell}(k,r)=f_1 $ is the first solution 
to the generalized spheroidal radial wave equation [12,14], and regular 
function $ g_{w_+,w_-}^{\ell}(k,r)=g_1 $ is the second one to the same 
equation, while $ S_{m,0}^{\ell}(ka,\theta) $ is an ordinary spheroidal 
angular wave function [13]. These functions can be orthonormalized to 
constitute their corresponding orthogonal complete functions [12,13,14]. 
In addition, scalar wave function $ \Phi $ has asymptotic behaviors of 
plane waves at infinity: 
$$ \Phi(t,r,\theta,\varphi) \rightarrow e^{i(\pm kr-\omega t+m\varphi)}
   S_{m,0}^{\ell}(ka,\theta), \hskip 1cm ( r\rightarrow \pm \infty )  $$

In fact, components $ \Phi_{r>r_+}^{\rm{out}} $ and $ \Phi_{r<r_+}^{\rm{out}} $
have asymptotic behaviors when $ r \rightarrow r_+ $
\begin{eqnarray}
 \Phi_{r>r_+}^{\rm{out}} \rightarrow c_1(r-r_+)^{iw_+}S_{m,0}^{\ell}(ka,
  \theta)e^{i(m\varphi-\omega t)},\hskip 0.5cm ( r \rightarrow r_+ ) \\ 
 \Phi_{r<r_+}^{\rm{out}} \rightarrow c_2(r-r_+)^{-iw_+}S_{m,0}^{\ell}(ka,
  \theta)e^{i(m\varphi-\omega t)},\hskip 0.5cm ( r \rightarrow r_+ )
\end{eqnarray}

Clearly, the outgoing wave $ \Phi_{r>r_+}^{\rm{out}} $ can't be directly 
extended from $ r_+ < r < \infty $ to $ r_- < r < r_+ $, but it can be 
analytically continued to an outgoing wave $ \Phi_{r<r_+}^{\rm{out}} $ 
that inside event horizon $ {\cal{H}}_+ $ by the lower half complex 
$ r $-plane around unit circle $ r=r_+-i0 $:
$$ r-r_+ \rightarrow (r_+-r)e^{-i\pi}. $$

By this analytical treatment, we have
\begin{equation}
 \Phi_{r<r_+}^{\rm{out}}(t,r,\theta,\varphi)
 \sim c_2(r-r_+)^{-iw_+}(r-r_-)^{iw_-}f_{w_+,w_-}^{\ell}(k,r)
 S_{m,0}^{\ell}(ka,\theta)e^{i(m\varphi-\omega t)}
\end{equation}

\noindent
here function $ f_{w_+,w_-}^{\ell}(k,r) $ can be analytically continued to
be function $ g_{w_+,w_-}^{\ell}(k,r) $.

As a difference factor $ (r-r_+)^{-2iw_+} $ emerges between functions 
$ F_{w_+,w_-}^{\ell}(k,r)=(r-r_+)^{iw_+}(r-r_-)^{iw_-}f_1 $ and 
$ G_{w_+,w_-}^{\ell}(k,r)=(r-r_+)^{-iw_+}(r-r_-)^{iw_-}g_1 $, then 
$ \Phi_{r>r_+}^{\rm{out}} $ differs $ \Phi_{r<r_+}^{\rm{out}} $ by a 
factor $ e^{2\pi w_+} $, thus we can derive a relation
\begin{equation}
 |\frac{\Phi_{r>r_+}^{\rm{out}}}{\Phi_{r<r_+}^{\rm{out}}}|^2=e^{-4\pi w_+}.
\end{equation}

Using the method of Damour-Ruffini's, it is easy to obtain an "external" 
thermal radiation spectrum [15]:
\begin{equation}
  \langle N_+ \rangle=|C_+|^2=\frac{1}{e^{4\pi w_+}-1}.
\end{equation}

Similarly, due to symmetry between exterior horizon and interior horizon, we 
can also establish an "internal" black body spectrum:
\begin{equation}
 \langle N_- \rangle=|C_-|^2=\frac{1}{e^{4\pi w_-}-1}
\end{equation}

\noindent
for a "right" outgoing wave $ \Psi^{\rm{out}}=\Psi^{\rm out}(t,r,\theta,
\varphi) $ from horizon $ {\cal{H}}_- $ which is similar to the above-head 
"left" outgoing wave $ \Phi^{\rm{out}} $,
\begin{equation}
 \Psi^{\rm{out}}=C_-[\eta(r_--r)\Psi_{r<r_-}^{\rm{out}}
 +\eta(r-r_-)\Psi_{r>r_-}^{\rm{out}}e^{2\pi w_-}]
\end{equation}

\noindent
and
\begin{equation}
 |\frac{\Psi_{r<r_-}^{\rm{out}}}{\Psi_{r>r_-}^{\rm{out}}}|^2=e^{-4\pi w_-}.
\end{equation}

\noindent
Now, we have made analytical extension for $ \Psi^{\rm{out}} $ by the upper 
half complex $ r $-plane around unit circle $ r=r_-+i0 $:
$$ r_--r \rightarrow (r-r_-)e^{-i\pi}. $$ 

\noindent
{\large 3. Generalized First Laws of Black Hole Thermodynamics: 
    Classical and Quantum }

It is convenient to introduce formally the following notations:
\[
\begin{array}{ll}
{\rm Reduced \hskip 3pt horizon \hskip 3pt area}: {\cal{A}}_{\pm}=r_{\pm}^2+a^2,& 
{\rm Horizon}: r_{\pm}=M\pm \varepsilon, \\
{\rm Surface  \hskip 3pt gravity}: \kappa_{\pm}=\frac{r_{\pm}-M}{{\cal{A}}_{\pm}}
=\frac{\pm \varepsilon}{{\cal{A}}_{\pm}}, &
{\rm Angular  \hskip 3pt velocity}: \Omega_{\pm}=\frac{a}{{\cal{A}}_{\pm}},\\
{\rm Electric \hskip 3pt potential}: \Phi_{\pm}=\frac{er_{\pm}}{{\cal{A}}_{\pm}}, &
{\rm Frequency}: \omega_{\pm}=m\Omega_{\pm}+q\Phi_{\pm}. 
\end{array}
\]

We can derive algebraically the following generalized first laws of 
classical and quantum black hole thermodynamics in both integral and  
differential forms on exterior horizon $ {\cal{H}}_+ $ as well as on 
interior horizon $ {\cal{H}}_- $ ( see Appendix ).
 
a. \underline{Generalized first laws of classical thermodynamics } in 
differential and integral forms [3,16]:
\begin{eqnarray}
 dM&=&\frac{\kappa_{\pm}}{2}d{\cal{A}}_{\pm}+\Omega_{\pm}dJ+\Phi_{\pm}de,\\
  M&=&\kappa_{\pm}{\cal{A}}_{\pm}+2J\Omega_{\pm}+e\Phi_{\pm}.
\end{eqnarray}

b. \underline{Generalized first laws of quantum thermodynamics } in integral 
and differential forms [9,17]:
\begin{eqnarray}
  \omega&=&2\kappa_{\pm}w_{\pm}+m\Omega_{\pm}+q\Phi_{\pm}, \\
 d\omega&=&2\kappa_{\pm}dw_{\pm}+\Omega_{\pm}dm+\Phi_{\pm}dq.
\end{eqnarray}

Relations (20, 21) demonstrate that electro-magnetic energy $ e\Phi_{
\pm} $ is an interaction energy (gauge term), while term $ \kappa_{\pm}
{\cal{A}}_{\pm} $ and term $ J\Omega_{\pm} $ being self energy terms.

\noindent
{\large 4. Quantum Conservation Laws }

Let us consider a complex scalar field $ \Phi $ in thermal equilibrium with
a Kerr-Newman black hole at a pair of local temperature $ {\cal{T}}_{\pm}=
\kappa_{\pm}/2 $. In thermal equilibrium radiation process, surface gravity 
, angular velocity and electrical potential can be considered to undertake 
little change. In virtue  of conditions that thermodynamical equilibrium 
could exist on horizons:
$$ \kappa_{\pm-0}=\kappa_{\pm+0},
   \Omega_{\pm-0}=\Omega_{\pm+0},
   \Phi  _{\pm-0}=\Phi  _{\pm+0}, $$

\noindent
combining differential relations Eq.(20) with Eq.(23), we can deduce 
five quantum conservation laws for energy, angular momentum, charge and 
entropy respectively:
\begin{eqnarray} 
  dM &=& nd\omega, ({\rm Energy}) \\
  dJ &=& ndm, ({\rm Angular \hskip 4pt Momentum}) \\
  de &=& ndq, ({\rm Charge})  \\
  \frac{1}{4}d{\cal{A}}_{\pm} &=& ndw_{\pm}, ({\rm Entropy}). 
\end{eqnarray}

\noindent
Here $ n $ is an integral multiplier. From integral relations (21) and 
(22), we can also obtain a special quantum state $ nm=J, n\omega=M/2, 
nq=e/2, nw_{\pm}={\cal{A}}_{\pm}/4. $

Eqs.(24-27) indicate that a Kerr-Newman black hole has discrete increment 
of energy, angular momentum, charge and entropy. When a KNBH is in dynamical 
equilibrium with a scalar field, it radiates the same quantity of quanta as 
that it absorbs. Thus the total quantities of energy, charge, entropy and 
angular momentum of the whole system being consisted of black hole and scalar
field quanta remain unchanged in this thermodynamical equilibrium radiation 
process [9].

\noindent
{\large 5. Entropy: Classical and Quantum }

In fact, Eq.(27) is a pair of generalized second thermodynamic laws in 
quantum form. By integrating this equation, we obtain quantum black hole 
entropy:
\begin{equation}
   nw_{\pm}=\frac{1}{4}{\cal{A}}_{\pm}+C_{\pm}.
\end{equation} 

As Bekenstein-Hawking's classical entropy [1,2] is $ S_{\pm}=A_{\pm}/4
=\pi{\cal{A}}_{\pm} $, quantum entropy $ nw_{\pm} $ are equivalent to 
the reduced entropy $ {\cal{S}}_{\pm}=S_{\pm}/(4\pi) $, so we have 
Bekenstein-Hawking relations (Choose constant $ C_{\pm}=0 $):
\begin{equation}      
 {\cal{S}}_{\pm}=nw_{\pm}={\cal{A}}_{\pm}/4. 
\end{equation}

Eqs.(28, 29) demonstrate that Bekenstein-Hawking classical entropy origins 
statistically from quantum entropy of quantized field, that is, the classical 
entropy of black hole is equal to quantum entropy of field [9].

\noindent 
{\large 6. Spectrum: Continued or Discrete? }

Quantum numbers of entropy $ nw_{\pm} $ must be integers as quantities $ A, B, 
\varepsilon $ correspond to angular momentum $ -m, -s, -a $ respectively when 
mass $ \mu=0$. It is suggested that the quantum entropy $nw_{\pm}$ be discrete 
numbers, namely be integers. The thermal spectrum $ \langle N_{\pm} \rangle $ 
for bound states are discrete spectrum, while the spectrum for scattering 
states being continual ones.

\noindent 
{\large 7. Temperature: Positive or Negative? }

From thermal spectrum of Hawking radiation:
$$ \langle N_{\pm} \rangle=\frac{1}{e^{4\pi w_{\pm}}-1}, $$
$$ w_{\pm}=\frac{\omega-m\Omega_{\pm}-q\Phi_{\pm}}{2\kappa_{\pm}} $$

\noindent
we can deduce that a KNBH has a pair of local temperature $ T_{\pm}=\kappa_{
\pm}/(2\pi)={\cal{T}}_{\pm}/\pi $ on horizons $ {\cal{H}}_{\pm} $.

If we accept the temperature interpretation of surface gravity $ \kappa_{
\pm}=\pm\varepsilon/(r_{\pm}^2+a^2) $, then temperature $ T_+ $ is positive 
while $ T_- $ being negative. The definition of negative temperature 
has no contradict with black hole having negative specific heat.

\noindent
{\large 8. Generalized Four Thermodynamical Laws} 

We give the chief points of four generalized laws of black hole 
thermodynamics as follows:

The Zeroth Law: The surface gravity $ \kappa_{\pm} $ of a stationary 
black hole (at equilibrium) are two constants on the entire surface
of its corresponding horizons $ {\cal{H}}_{\pm} $.

The First Law: In an isolated system including black holes, the total energy
of the system is conserved. 

The Second Law: The total entropy $ S_{\rm T}=S_{\rm BH}+S_{\rm M} $, never 
decreases in any physical process, where $ S_{\rm M} $ is the total entropy 
of ordinary matter outside black holes, $ \delta S_{\rm T} \geq 0. $ 

The Third Law: It is impossible by any physical process to reduce $ 
\kappa_{\pm} $ to zero by a finite sequence of operations. However, this
can be violated by quantum vacuum fluctuations. Quantum evaporation effect
can make a KNBH undertake a second order phase transition [18] from the 
non-extreme case ($ M^2\not=a^2+e^2 $) to the extreme case ($ M^2=a^2+e^2 $). 

Quantum Conservation Laws: In a thermal equilibrium process of black hole 
radiation, the total energy, total charge, total angular momentum and total
entropy of the whole system are conserved. 

To summarize, many results on the exterior horizon are generalized to similar 
ones on the inner horizon. We suggest that there should exist an interior 
radiation on the Cauchy horizon provided that the whole spacetime is described 
by the Kerr-Newman line element. This provides a rather good interpretation 
of the origin of black hole classical entropy arising statistically from 
quantum entropy of field quanta. Thus, if a KNBH really has two pairs of 
temperature and entropy, then how do we interpret them ? There exists a 
proposal that a KNBH be a two-energy levels system endowed with a pair of 
local temperature on the horizons. As far as existed theories are concerned, 
it seems to have no reason to exclude a negative temperature $ \rm{T}_-$.

\noindent {\bf Acknowledgment}

This work is supported partly by the NSFC and Hubei-NSF in China.

\vskip 0.5cm
\noindent
{\large APPENDIX: Algebraical Derivation of First Laws of Thermodynamics 
in Classical and Quantum Forms } 

Derivation: By differentiating equality $ {\cal{A}}_{\pm}=r_{\pm}^2+a^2=2M
r_{\pm}-e^2 $, we can deduce a relation $ r_{\pm}dM=\pm\varepsilon dr_{\pm}
+ada+ede $. Then multiplying this formula by $ r_{\pm} $ and adding a term 
$ a^2dM $, we obtain a relation $ {\cal{A}}_{\pm}dM=\pm\varepsilon/2
d{\cal{A}}_{\pm}+adJ+er_{\pm}de $. Eq.(20) is obtained by dividing this
relation with $ {\cal{A}}_{\pm} $.
 
From $ r_{\mp}({\cal{A}}_{\pm}+e^2)=2Mr_+r_-=2M(a^2+e^2)=2Ja+2Me^2 $, we can
deduce a relation $ (M\mp\varepsilon){\cal{A}}_{\pm}=2Ja+(2M-r_{\mp})e^2=2Ja
+e^2r_{\pm} $. Then Eq.(21) is obtained by dividing this equality with
$ {\cal{A}}_{\pm} $ and replacing terms $ \kappa_{\pm}{\cal{A}}_{\pm}=\pm
\varepsilon $.

From equalities 
$$ A=(\omega-\omega_+)/(2\kappa_+)-(\omega-\omega_-)/(2\kappa_-), 
B=(\omega-\omega_+)/(2\kappa_+)+(\omega-\omega_-)/(2\kappa_-) $$

\noindent 
we can obtain relations $ w_{\pm}=(B\pm A)/2=(\omega-\omega_{\pm})/(2
\kappa_{\pm}) $. Eq.(22) is obtained by multiplicating this relations 
with $ 2\kappa_{\pm} $ and displacing terms $ \omega_{\pm}=m\Omega_{\pm}
+q\Phi_{\pm} $. Then by differentiating this equation, we obtain
Eq.(23).

\end{document}